\documentstyle[12pt,aaspp4]{article}
\begin{document}
\def\um		{$\mu$m}
\def\i		{{\sc~i}}
\def\ii		{{\sc~ii}}
\def\etal	{{et al.}}
\slugcomment{Submitted The Astrophysical Journal: \today }

\title{{\bf{\it COBE}} FIRAS observations of Galactic Lines}
\author{
D. J. Fixsen\altaffilmark{1}\altaffilmark{3}, 
C. L. Bennett\altaffilmark{2}, and  
J. C. Mather\altaffilmark{2}}
%R. A. Shafer\altaffilmark{2}}

\altaffiltext{1}{Raytheon-STX Corp., Code 685, NASA/GSFC, Greenbelt, MD 20771}
\altaffiltext{2}{Code 685, Infrared Astrophysics Branch, Goddard Space Flight 
Center, Greenbelt, MD 20771}
\altaffiltext{3}{email address fixsen@stars.gsfc.nasa.gov}

\begin{abstract}

The {\it COBE} Far InfraRed Absolute Spectrophotometer (FIRAS) observations
constitute an unbiased survey over the wavelength range from 100 \um\ to 1 cm 
over 99\% of the sky. Improved calibration of the FIRAS instrument and the 
inclusion of all of the FIRAS data allow an improved signal to noise 
determination of the spectral lines by a factor of $\sim2$ over our previous 
results. The resolution is low (0.45 cm$^{-1}$) so only the strongest lines 
are observable. The CO chain from $J=1-0$ to $J=8-7$ is observed towards the 
Galactic center. The line ratios are roughly consistent with a 40~K 
excitation temperature. The 157.7 \um\ C\ii\ and 205.3 \um\ N\ii\ 
lines are observable over most of the sky. The 370.4 \um\ and 609.1 
\um\ lines of C\i, and the 121.9 \um\ line of N\ii\ are observed in the Galactic
plane.  The line ratios at the Galactic center are consistent with a 
density of $n_0\sim30$~cm$^{-3}$ and a UV flux of 
$G_0\approx15~\mu$W~m$^{-2}$~sr$^{-1}$ (10 Habing units).
The 269 \um\ H$_2$O line is observed toward the Galactic center in absorption.

\end{abstract}

\keywords{ISM: atoms, bubbles, general, molecules --- line: identification
Galaxy: general}

\section{Introduction}

An unbiased far infrared survey of the spectral line emission from the Galaxy 
has been reported from the Cosmic Background Explorer ({\it COBE}) mission by 
Wright \etal\ (1991) and Bennett \etal\ (1994). A new analysis of the FIRAS 
data provides an improvement of approximately a factor of 2 in signal-to-noise 
and a more complete map of the sky.  This has enabled improved full sky maps 
of the 157.7 \um\ C\ii\ and 205.3 \um\ N\ii\ cooling lines and additional 
detections of CO lines, and the 115.8 \um\ CH line. The 269 \um\ H$_2$O line was 
detected in absorption towards the Galactic center.

While the primary purpose of the instrument was to make precise measurements 
of the cosmic microwave background radiation, the instrument design had 
many advantages for the study of large scale diffuse emission from our
Galaxy.  The FIRAS has made absolute measurements of intensity 
with no beam switching, with a 7$^\circ$ beam, and has observed 99\% of the sky. 
This makes the FIRAS data well-suited for determination of diffuse emission.  
The FIRAS spectral resolution is low, as discussed below in \S 3, and no 
spectral lines are resolved.  Despite this, the Galactic rotation can 
be discerned using the detected FIRAS spectral lines (Fixsen \etal\ 1996).

The absolute nature of the FIRAS measurements coupled with the nearly full
sky coverage allow observation of the overall Galactic energy balance.
However care needs to be taken in comparing the results to measurements
taken with a small beam or done with beam switching.  Since the FIRAS beam is
large, even bright small sources can appear insignificant while large
diffuse regions that don't even appear in beam switched studies dominate.

The first cosmological results from the {\it COBE} FIRAS instrument were 
reported by Mather \etal\ (1990), and Boggess \etal\ (1992). More recent results 
are reported by Bennett \etal\ (1994), Fixsen \etal\ (1996, 1997, 1998, 1999),
and Dwek \etal\ (1997). Previous results of the {\it COBE} FIRAS unbiased far 
infrared survey of the Milky Way were published by Wright \etal\ (1991) and
Bennett \etal\ (1994).  Bennett's Figure 2 summarizes spectral line detections 
from {\it COBE} FIRAS: 157.7 \um\ C\ii, 370.4 \um\ and 609.1 \um\ C\i, J=2-1 
through J=5-4 CO, and 121.9 \um\ and 205.3 \um\ N\ii. Since then, improved data 
processing algorithms and techniques have allowed for a more 
detailed examination of the data.

The C\ii, C\i\ and CO emission lines detected by {\it COBE} are recognized 
as the expected cooling lines from PDRs (Hollenbach, Takahashi, \& Tielens 1991).
The result that the Galactic-averaged PDR properties are described by lower 
excitation conditions than objects such as Orion is reinforced by the 
weakness of the 145.5 \um\ O\i\ line.  

\section{Calibration and Line Fitting}

The FIRAS instrument and its calibration were described in detail by Fixsen 
\etal\ (1994).  Significant improvements were noted by Fixsen \etal\ (1996). 
In this paper we use the combined (all detectors and all scan modes), destriped 
data with higher frequency resolution also denoted ``Pass 4" (FIRAS Explanatory
Supplement). The improved calibration, destriping, and larger data set reduce
the uncertainties to about half of those of Bennett \etal\ (1994).
Data from all of the channels, scan modes, and frequency bands were combined 
using a weighted average to form a data set with 210 spectral points 
from 4.4 mm to 104 \um\ (2.3 to 96.3 cm$^{-1}$).  Some of these data were 
used by Bennett \etal\ (1994).

The instrument line response function is described in detail by Bennett \etal\
(1994). Since the apodization is asymmetric, the resulting line profile is 
complex. For ``Pass 4" we padded the 512 point interferograms 
with an additional 128 points on the short side to include almost all of the 
information in the real part of the spectrum with a bin width of 0.45 cm$^{-1}$.

Continuum and line spectra were fitted to the data as follows. First, a 2.728 K 
%(the 3 mK correction from Fixsen 1999 is not in this data set) 
Planck spectrum, $P_\nu(T=2.728)$ was subtracted from the spectrum at each sky 
pixel. Then the spectrum at each pixel was fit with 36 parameters. The parameters 
of the fit (and their corresponding degrees of freedom at each pixel) were: 
the amplitudes at each assumed line frequency (18), the amplitude of a 
$dP/dT$ term to remove the cosmic microwave dipole (1), the amplitude of 
a $\nu^{1.6}*P_\nu(18)$ to remove most of the Galactic dust emission (1), and 
coefficients of Legendre polynomials for the continuum emission of the dust (16).  
The fit removes 36 degrees of freedom, leaving 174 degrees of freedom at each 
of 6063 pixels.  The data were weighted by an inverse variance that is dominated 
by detector noise, but calculation of the uncertainties include other terms.
The use of this diagonal matrix weighting is an approximation to the
full covariance matrix, which is not perfectly diagonal because of the
effects of the apodization and the calibration process. The full covariance
matrix was used to compute the uncertainties (Fixsen \etal\ 1994).

The Galactic plane is unresolved by the 7$^\circ$ FIRAS beam. From DIRBE and 
other observations it is seen that for (1~cm$>\lambda> 100$~\um) the 
Galactic radiation is concentrated in a band 1$^\circ$ wide. For the Galactic 
plane we use the FIRAS data before they are binned into pixels.
We form bins of 5$^\circ$ Galactic longitude and 1$^\circ$ Galactic latitude 
centered on the Galactic plane. With the assumption that the observations 
near the Galactic plane are dominated by the radiation from a region 
1$^\circ$ across at the Galactic plane, the spectrum for each
bin is given by:
\begin{equation}
S_\nu=\sum_i (W_i B_i S_{i\nu})/\sum_i(W_i B_i^2),
\end{equation}
where $W_i$ is the weight of the $i^{th}$ data in the bin, $B_i$ is the
beam response to a line source at the Galactic plane and $S_{i\nu}$ is the
$i^{th}$ spectrum. The beam response is derived from observations near the moon,
which are not otherwise used.  The beam response is used to derive $B$, which
is normalized to a 1$^\circ$ wide unit
line source.  This has two important effects. First we increase the signal to
noise ratio in the result as we weight the data in proportion to the signal.
Second the amplitude is increased by a factor of $\sim8$ over the presentation
in Bennett \etal\ (1994).  This is because Bennett \etal\ give an average over 
$|b| < 5^\circ$ while we give the result for $|b| < 0.^\circ 5$.
For $|b| > 10^\circ$ we use the pixels from Fixsen (1996).  

For each pixel we perform a simultaneous least squares fit 
of a continuum spectral baseline plus a series of spectral line 
profiles centered on the wavelengths of known lines.  The result is a set of
spectral line intensities for each pixel.  We find that while the 
C\ii\ emission (Figure 1, top) follows the continuum closely, and 
the N\ii\ emission (Figure 1, bottom) follows it 
approximately, the CO and other emissions have distinct distributions.

\section{Analysis Results}

\subsection{The CO Emission}

The J=1-0, 6-5, 7-6 and 8-7 CO lines are detected, as well as the CO J=2-1, 
3-2, 4-3 and 5-4 transitions noted by Bennett \etal\ (1994). Figure 3 shows
CO emission from the central Galactic region. The 7-6 
transition appears but the 370.4 \um\ C\i\ and the 369 \um\ 7-6 CO transitions 
are too close to be resolved by the FIRAS. Some care must be exercised to 
interpret Figure 3g. This is probably a mixture of the 7-6 CO transition and 
the C\i\ emission. 

Bally \etal\ (1988) found that in the Galactic center region the low lying 
rotational CO lines
are optically very thick. So, the line emission should be proportional to 
$\nu^4*exp(-E/kT)$, where the extra factor of $\nu$ is because the same 
velocity dispersion leads to higher bandwidth at higher frequencies. With this 
model, the full CO spectrum can be reasonably approximated by a 40 K 
temperature (figure 2a). A simple explanation is a temperature distribution 
with a peak at 40~K. We use this model to estimate the emission due to CO 7-6 
and assign the remaining emission to the 370.4 \um\ C\i. 

There are small regions exhibiting much higher temperatures (Simpson \etal\ 
1997, Jackson \etal\ 1996). We see only a hint in these data in the
form of the high CO 8-7 transition and the uncertainties do not justify
a more complex model for the large area-averaged data. The larger 
uncertainties at higher frequencies allow
only upper limits of 10 and 15~nW~m$^{-2}$~sr$^{-1}$ respectively for the 
CO 9-8 and CO 10-9 lines. These results are in broad agreement with Paglione
\etal\ (1998) who concluded 80\% of the gas is 40 K and below.  Since the 
CO is optically thick significant high temperature regions could be hidden
from the FIRAS instrument.

There is also a smaller but significant amount of CO emission in the plane at 
longitudes $2.5<|l|<32.5^\circ$. Again we use the thick optical model of CO to 
fit to a 22 K spectrum (Figure 2b), estimate the emission due to CO 7-6 and 
assign the remaining emission to the 370.4 \um\ C\i. These are used in the 
table.

Over most of figure 2g the signal to noise is too low to estimate the 
CO temperature. Instead the CO 7-6 emission is approximated as the 
average of the CO 8-7, CO 6-5 and CO 5-4 emission.  The resulting C\i\ (Figure 
5c) shows a similar intensity and structure to the 609.1 \um C\i\ emission.

\subsection{Water Absorption}

The ground state $1_{11}-0_{00}$ 269 \um\ transition of para-H$_2$O is detected.
The longitude profile of this water transition, 
shown in Figure 4a, displays an absorption dip towards the Galactic center, 
as do Bennett \etal\ (1994). With the improved signal to noise ratio we can now 
detect this line at the $\sim5\sigma$ level.  
The error bars on the line intensity are larger near the Galactic center
because of uncertainty in the ability to calibrate and remove the larger
continuum emission there. As shown by Bennett \etal\ (1994) the water absorption
should be expected.

%Water has a dipole moment of $\mu_D=1.9$ Debye, corresponding to the 269 \um\ 
%para-water transition rate $A=1.86\times 10^{-2}$ s$^{-1}$. If we assume the 
%abundance of water is [H$_2$O]/[H$_2$]$=10^{-6}$, that 10\% of all water 
%resides in the gaseous ground state, and that the density of an absorbing 
%cloud is typically $10^2$ cm$^{-3}$, then the spectral line will be optically 
%thick.  Hence, we expect optically thick water absorption in this transition 
%to be common. For a linewidth corresponding to a velocity dispersion of 30 km 
%s$^{-1}$ (a typical width for lines near the GC), the FIRAS observation would 
%suffer a frequency dilution factor of 120. Taking the mean value of the 
%negative intensity dip as a real absorption gives $\Delta I=-15$ 
%nW~m$^{-2}$~sr$^{-1}$ with a continuum intensity at that location and 
%wavelength of $I=1250$ nW~m$^{-2}$~sr$^{-1}$, so $\Delta I/I=1.2\times 10^{-2}$. 
%Correcting for the frequency filling factor gives
%$120 \Delta I/I\approx 1$, so the measurement is consistent with optically thick
%water absorption.  The corresponding ground state transition of ortho-water
%is at 538 \um\ (Figure 4b), also observable by  FIRAS.  This transition will 
%also be optically thick, so any absorption feature in 269 \um\ para-water should
%appear at 538 \um\ at approximately a fifth of the 269 \um\ intensity.
%The uncertainty of the signal at 538 \um\ does not allow detection of a signal
%this size. 

In Figure 4c we examine the ratio of the 269 \um\ para-H$_2$O 
line to the continuum. The change in the ratio shows that this is a true
absorption and not an artifact of the calibration. Beyond the central
Galactic region ($40^\circ>l>-90^\circ$) the signal is so small that it is swamped
by the noise.

\subsection{Emission Lines of O, Si, and CH}

As oxygen is the third most abundant element, molecular oxygen might be expected
in a wide variety of conditions (Viala \& Walmsley 1996). Measuring it
would help to unravel the many reactions involving oxygen. Searching for it
is difficult because of the oxygen in the atmosphere. The ($3_2-1_2$) 
transition of O$_2$ at 705.8 \um\ (Figure 4e) is predicted by Marechal \etal 
(1997) to be among the strongest O$_2$ lines. The intensity could be up to 10 
nW~m$^{-2}$~sr$^{-1}$ for thick clouds. The line was not detected by {\it COBE} 
FIRAS, and we place a limit (95\% CL) on its intensity towards the Galactic 
center of 1 nW~m$^{-2}$~sr$^{-1}$. There is 1 nW~m$^{-2}$~sr$^{-1}$ of O$_2$ 
in a ring ($|l|\approx10\deg$) but this is only a $2.5\sigma$ effect and we do 
not claim a detection, but only an upper limit of 3 nW~m$^{-2}$~sr$^{-1}$.

The 145.5 \um\ O\i\ (Figure 4f) is detected but 
it is strongest in the spiral arms rather than the Galactic center. The 115.8 
\um\ CH (Figure 4h) line has been detected, although the FIRAS noise is high. 
A weighted average using the 205.3 \um\ N\ii\ intensity as a template improves 
the signal/noise ratio to show the detection of the 115.8 \um\ CH line at 
the $\sim 3 \sigma$ level. Using the same method, the 129.7 \um\ Si\i\ line 
shows a $\sim 2 \sigma$ effect (Figure 4g), but we cannot claim a detection
and we place an upper limit (95\% CL) of 250 nW~m$^{-2}$~sr$^{-1}$ towards the 
Galactic center. None of these lines are evident out of the Galactic plane. 

Liu \etal\ (1997) made a careful search for CH lines in the planetary nebula 
NGC 7027.  They report detections of the rotational lines 
$^2\Pi_{1/2}($F$_2)J=3/2-1/2$ at 149.2 \um\ (4.6 nW~m$^{-2}$~sr$^{-1}$), 
the  $^2\Pi_{1/2}($F$_2)J=5/2-3/2$ lines at 180.7 \um\ (3.7 nW~m$^{-2}$~sr$^{-1}$) 
and the $^2\Pi_{3/2}($F$_2)J=5/2-3/2$ lines at 115.8 \um\ (2.2 nW~m$^{-2}$~sr$^{-1}$). 
They also report an upper limit of 1 nW~m$^{-2}$~sr$^{-1}$
on the $^2\Pi_{1/2}($F$_2)J=7/2-5/2$ lines at 118.6 \um. The intensities given
here are the mean intensities for the 70 arc second beam of the {\it ISO}
long wavelength spectrometer.  Most of the radiation originates from a region
$\sim5$ arc seconds across (Middlemass 1990), which implies an intensity of 
440 nW~m$^{-2}$~sr$^{-1}$ for the central region. The intensity we observe 
towards the Galactic center region (that is the region weighted by the N\ii) 
is $\sim70$ nW~m$^{-2}$~sr$^{-1}$. Liu \etal\ (1996) observe the 157.7 \um\ C\ii\ 
to 115.8 \um\ CH ratio to be about 150. The ratio we observe using the same 
selection for C\ii\ as for the CH is about 300. The uncertainties are large
so the difference may be insignificant or the mean inner Galaxy may have a
higher C\ii\ to CH ratio than NGC 7027.

The first excited rotational level of CH has a critical density of 
$\sim4\times10^6$ cm$^{-3}$ and the next level has a critical density of 
$\sim10^8$ cm$^{-3}$ (Phelps \& Dalby 1966). Although we do not see the 
149.2 \um\ lines they are in the wings of the very strong 157.7 \um\ C\ii\ 
line and the upper limit would allow the same ratio as observed by Liu \etal 
(1997). Evidently at least 
part of the radiation emanates from fairly dense regions.

\subsection{C\ii, C\i}

The 157.7 \um\ C\ii\ line is clearly the dominant line in this entire band.
Examination of the intensity along the Galactic plane (Figure 5a) shows
the peak of the C\ii\ line is not in the Galactic center but in the molecular
ring at $l=\pm25^\circ$. Other peaks are evident at $\pm50^\circ, 80^\circ, 
110^\circ$, and $135^\circ$,
with the inner peaks more evident for $l<0$ and the outer peaks more evident
for $l>0$, clearly demonstrating the spiral structure of the Galaxy.

The C\ii\ to continuum ratio is fairly constant both along the Galactic plane
and out of the plane with C\ii\ enhancements of a factor of 2 or 3 in the 
spiral arms (Figure 5c).
 
The 609.1 \um\ C\i\ (Figure 5b) is almost entirely confined to the inner 
Galactic region, with a clear peak at the Galactic center and shoulders out to 
the molecular ring, unlike the C\ii\ which has a dip in the Galactic center.

The 370.4 \um\ C\i\ line can not be separated from the 269 \um\ CO 7-6 
transition, with the FIRAS resolution. We model the CO 7-6 emission as the 
average of the CO 8-7, CO 6-5 and CO 5-4 emission.  The resulting C\i\ (Figure 
5c) shows a similar intensity and structure to the 609.1 \um\ C\i\ emission.

\subsection{N\ii}

The 205.3 \um\ N\ii\ line is strong enough to form a reasonable full sky map 
from the FIRAS data (Figure 1 bottom). It has a similar structure to the C\ii\ 
(Figure 5e) although it is more strongly concentrated in the Galactic center. 
The N\ii/C\ii\ ratio plotted as a function of Galactic longitude (Figure 5h) 
shows a smooth function with a peak at the Galactic center.
Rudolph \etal\ (1997) give an N/S abundance ratio variation as 
$\propto 10^{-.03R}$, where $R$ is the radius in kpc. They note that the S/O 
ratio is constant as expected. Assuming the S/C ratio is constant, we derive 
the dashed line in figure 5h for radiation dominated by the inner edge,
or the dotted line for uniform sources.

The 121.9 \um\ N\ii\ line (Figure 5f) has a structure and magnitude similar to 
the 205.3 \um\ N\ii\ line emission. The higher noise of the FIRAS at this 
frequency prevents a detailed comparison but to first order it appears that 
the 121.9 \um\ N\ii\ line emission can be modeled as proportional to the 
205.3 \um\ N\ii\ line emission. The mean of the $I$(121.9 \um\ N\ii)
/$I$(205.3 \um\ N\ii) ratio is $1.1\pm0.1$ suggesting most of the N\ii\ 
radiation originates in low density regions (Rubin 1985).  

\clearpage
\begin{table}[tbph]
\begin{center}
\caption{Line Flux at High Galactic Latitude}
\begin{tabular}{lrrrr}
\tablewidth{7.5in}
   & Galactic Center & Inner Galaxy & Outer Galaxy & High Latitudes \\
Line  & $|l|<2.5^\circ$ & $2.5<|l|<32.5^\circ$ & $|l|>32.5^\circ$ & $|b|>10^\circ$ \\
\hline
CO 1-0 & $  1.6\pm0.5$ & $ 0.5\pm0.3$ & $ 0.2\pm0.2$ & $ 0\pm.01$ \\
CO 2-1 & $  6.4\pm0.3$ & $ 2.3\pm0.2$ & $ 0.5\pm0.1$ & $ 0\pm.01$ \\
CO 3-2 & $ 11.8\pm0.5$ & $ 3.8\pm0.3$ & $ 0.7\pm0.2$ & $ 0\pm.01$ \\
CO 4-3 & $ 17.7\pm0.6$ & $ 3.4\pm0.3$ & $ 0.5\pm0.3$ & $ 0\pm.01$ \\
CO 5-4 & $ 16.5\pm1.0$ & $ 2.9\pm0.6$ & $ 0.9\pm0.5$ & $.01\pm.01$ \\
CO 6-5 & $ 11.5\pm1.6$ & $ 0.5\pm1.0$ & $-0.2\pm0.7$ & $ 0\pm.01$ \\
CO 7-6$^1$ &$10\pm1.5$ & $ 0.3\pm1.0$ & $ 0.1\pm0.3$ & $ 0\pm.01$ \\
CO 8-7 & $ 10.8\pm1.4$ & $ 1.8\pm0.8$ & $ 0.1\pm0.5$ & $.01\pm.01$ \\

C\i\ 609 \um & $ 11\pm0.6$ &$  5\pm0.4$& $ 1.4\pm0.3$ & $ .01\pm.01$ \\
C\i\ 370 \um$^1$&$11\pm1.9$&$  7\pm1.0$& $ 1.4\pm0.5$ & $   0\pm.01$ \\
C\ii\ 158 \um& $875\pm32$ &$1021\pm17$ & $ 254\pm5 $ & $ 1.48\pm.07$ \\
N\ii\ 205 \um& $ 97\pm6 $ &$ 107\pm3 $ & $  18\pm1  $ & $ .05\pm.02$ \\
N\ii\ 122 \um& $ 76\pm51$ &$  23\pm22$ & $   2\pm9 $ & $ .17\pm.14$ \\
O\i\ 146 \um & $ 29\pm29$ &$  24\pm13$ & $   5\pm5  $ & $ .07\pm.08$ \\
CH 116 \um  & $149\pm82$ &$  14\pm34$ & $  15\pm15 $ & $ -.05\pm.25$ \\
Dust Emission & 130000 & 92000 & 25000 & 150 \\
\hline
\multicolumn{5}{l}{Units are nW m$^{-2}$ sr$^{-1}$. 
Uncertainties are 1$\sigma$ and include systematic effects.} \\
\multicolumn{5}{l}{$^1$The CO 7-6 was estimated from the other CO lines and
the residual was ascribed to C\i.} \\
\end{tabular}
\end{center}
\end{table}

\subsection{Galactic Center}
In figures 1, 3, 4, and 5 the Galactic center stands out as a distinct region.
Only there is the CO 1-0 detectable. The CO line ratios suggest a temperature 
of 40~K (Figure 2). We attempt to fit the observed line strengths of CO 1-0
thru CO 8-7, the C\i\ lines and the C\ii\ line by adjusting the density,
UV flux and an overall gain to the the PDR model of Hollenbach \etal\ 
The best fit is $n_0=30~$cm$^{-1}$ and $G_0=15~\mu$W~m$^{-2}$~sr$^{-1}$ 
(10 Habing units). The fit is
only weakly dependent on the density, allowing a factor of 5 in either direction.
The data constrain the UV flux much better, allowing only a factor of 2 in $G_0$.
The fit is poor ($\chi^2/$DOF=58) with the major problem being the observed
C\ii\ line intensity which is a factor of 3 higher than the model. This might be
due to foreground or background emission from the molecular ring.
Thus we can make the model agree with the ratios if we assume a density of 
$\sim30$~cm$^{-3}$ and a UV density of $G_0=15~\mu$W~m$^{-2}$~sr$^{-1}$ 
and that much of the observed C\ii\ flux at the Galactic center is really 
from the molecular ring. The density is in agreement with the Bally \etal\ (1988) 
average density estimate of 50 cm$^{-1}$ over the region $r<500$~pc, 
$|h|<100$~pc, which is approximately the same as our Galactic center region. 

The values we obtained for $G_0$ and $n_0$ are lower than other
observers (Simpson \etal\ 1997) have found in the Galactic Center.  They used
much smaller beamwidths and were sensitive to compact objects, while our
beamwidth of 7$^\circ$ corresponds to a region 1 kpc across.  Our numbers
are also small compared to those we found for the inner Galaxy region
defined below. We conclude that the high brightness of the Galactic Center
is due to relatively low density material filling a relatively large
fraction of available space, rather than compact objects around young stars.

\subsection{Inner Galaxy}
A second distinct region is the inner Galaxy, roughly the Galactic plane with
$2.5<|l|<32^\circ$. This region has much smaller CO emission even though it has
more C\i\ emission. The CO emission is consistent with an optically thick 
temperature of 22~K, considerably cooler than the Galactic center. Fitting the 
PDR model to the data yields $n_0>5000$~cm$^{-3}$ and $G_0>800~\mu$W~m$^{-2}$~sr$^{-1}$. 
Even so there is a factor of 2 more C\ii\ emission observed than the model 
prediction. This suggests that in the inner Galactic regions the lines 
emission is from dense clouds with high UV flux, and that the C\ii\ emission 
is more efficient than in the one dimensional PDR model. This could be a
simple surface to volume ratio effect.

\subsection{Outer Galaxy}
The outer Galactic plane presumably should also be similar to local conditions
averaged over a sufficiently large volume. This region has approximately the 
same line ratios as the inner Galaxy suggesting the conditions are much the
same on average although the average emission is a factor of 5 lower than the 
inner Galaxy. The one exception is that the N\ii\ to C\ii\ ratio is lower.  This
can be explained by a generally lower N abundance in the outer Galaxy
(Rudolph \etal 1997).

\subsection{High Latitude Emission}
The high latitude emission is much weaker than the emission in the outer 
Galactic plane.  The simple picture of a thin disk ($R/t\sim100$) gives just
this effect with the same line ratios as observed in the outer Galaxy.
Thus description holds although the signal/noise ratios for most of the lines
are too small to make a strong statement.

\section{Conclusions}

\noindent 1. The {\it COBE} FIRAS has conducted an unbiased survey of the 
far-infrared emission from our Galaxy.  Previous results of this survey were 
reported by Wright \etal\ (1991) and Bennett \etal\ (1994). We report here the 
distribution along the Galactic plane ($b=0^\circ$) of the spectral line
emission for the detected spectral lines.

\noindent 2. The CO line ratios suggest an excitation temperature of 40~K in 
the Galactic center and $\sim20$~K in other typical regions of CO emission.

\noindent 3. We find the PDR model of Hollenbach fits the observed line ratios 
in the Galactic center if we assume a density of $\sim30$~cm$^{-3}$ and a 
UV flux of $G_0\approx15~\mu$W$~$m$^{-2}~$sr$^{-1}$).

\noindent 4. We find the C\ii\ emission observed is higher than the PDR 
model of Hollenbach over the source regions in the rest of the Galaxy, even 
if we assume high densities $>5000$~cm$^{-3}$ and high UV flux 
$G_0>800~\mu$W$~$m$^{-2}~$sr$^{-1}$ in the emission regions.

\noindent 5. The 205.3 \um\ line of N\ii\ generally follows the
157.7 \um\ C\ii\ line distributions, but there is probably an abundance
effect that enhances the N\ii\ intensity in the inner Galaxy.

\noindent 6. We have detected the 115.8~\um\ CH line in the inner Galactic
region and the 269~\um\ H$_2$O line in absorption towards the Galactic center.

\acknowledgments

{\it COBE} is supported by NASA's Astrophysics Division.  
We are grateful for the contributions of the entire {\it COBE} Science Team 
and the support personnel at Goddard's Cosmology Data Analysis Center (CDAC).
We thank G. Hinshaw for help with plots and pictures.

\clearpage

\begin{center}
{\bf FIGURE CAPTIONS}
\end{center}

{\bf Figure 1}:(Color Plate)  
The maps are projections
of the full sky in Galactic coordinates.  The plane of the Milky Way is
horizontal in the middle of the map with the Galactic center at the center.
Galactic longitude $l=90^\circ$ is left of center.  The maps are smoothed
to 10$^\circ$ resolution. The black stripes are unobserved portions of sky.
Color bars indicate emission intensity in nW~m$^{-2}$~sr$^{-1}$. The scale 
is not linear. 
{\it (top)}: A full sky map of 157.7 \um\ C\ii\ emission from the {\it COBE} 
FIRAS experiment. 
{\it (bottom)}: A full sky map of 205.3 \um\ N\ii\ emission from the {\it COBE} 
FIRAS experiment.

{\bf Figure 2}:The spectrum of the CO chain at the Galactic center.  The solid
line is a 40 K fit. The dotted lines are 35~K and 45~K.
The CO 7-6 point was not used in the fit. The error bars are $1\sigma$.
The lower plot is the inner Galactic region with a fit of 22~K.

{\bf Figure 3}: 
Longitude profiles of the intensity for CO lines for the Galactic 
plane.  The intensity in each bin represents the average over $5^\circ$ 
in Galactic longitude and $\pm 0.5^\circ$ in Galactic latitude, under the 
assumption that the emission is dominated by an unresolved line at the Galactic
plane.  The 2$\sigma$ error bars include detector noise and gain uncertainties.
There are few measurements in the $l=90^\circ$ bin, so we caution against 
overinterpretation of results based only on this data point.
The transitions are noted on the plots.  Note the CO 7-6 transition plot 
also contains the C\i\ $2p^2:^3$P$_2-^3$P$_1$ transition at 370.4 \um.

{\bf Figure 4}: 
Plots are prepared as in figure 3 but the vertical scales are different.
(a) The detected (see text) para-H$_2$O $1_{11}-0_{00}$ ground state 
transition at 269.3 \um.
(b) The undetected ortho-H$_2$O $1_{10}-1_{01}$ ground state transition at 
538.3 \um.
(c) The ratio of the para-H$_2$O $1_{11}-0_{00}$ to the continuum showing
the fractional absorption (dimensionless scale).
(d) The total IR emission ($\lambda>105$ \um), the scale is 
$\mu$W~m$^{-2}$~sr$^{-1}$.
(e) The undetected O$_2$ $3_2-1_2$ at 705.8 \um.
(f) The detected O\i\ $2p^4:^3$P$_0-^3$P$_1$ transition at 145.5 \um.
(g) The undetected Si\i\ $3p^2:^3$P$_1-^3$P$_0$ transition at 129.7 \um.
(h) The detected CH transition at 115.8 \um.

{\bf Figure 5}: 
(a) The C\ii\ 2p:$^2$P$_{3/2}-^2$P$_{1/2}$ fine structure ground
state transition at 157.7 \um.  
(b) The C\i\ $2p^2:^3$P$_1-^3$P$_0$ transition at 609.1 \um.
(c) The C\i\ $2p^2:^3$P$_2-^3$P$_1$ transition at 370.4 \um. 
This is residual after subtracting an estimate of the CO 7-6 emission.
(d)The ratio of C\ii\ to total far infrared continuum emission at wavelengths 
longer than 105 \um\ (dimensionless scale).
(e) The N\ii\ $2p^2:^3$P$_1-^3$P$_0$ fine structure ground state 
transition at 205.3 \um.  
(f) The N\ii\ $2p^2:^3$P$_2-^3$P$_1$ transition at 121.9 \um.  
(g) The ratio of N\ii\ 122 \um\ to N\ii\ 205 \um\ emission for the central
Galactic region (dimensionless scale).  Farther from the Galactic 
center the ratio is dominated by noise.
(h) The N\ii\ 205 \um\ to C\ii\ 157 \um\ ratio (dimensionless scale).
The dotted and dashed lines are predictions from simple models (see text).

\clearpage
\eject

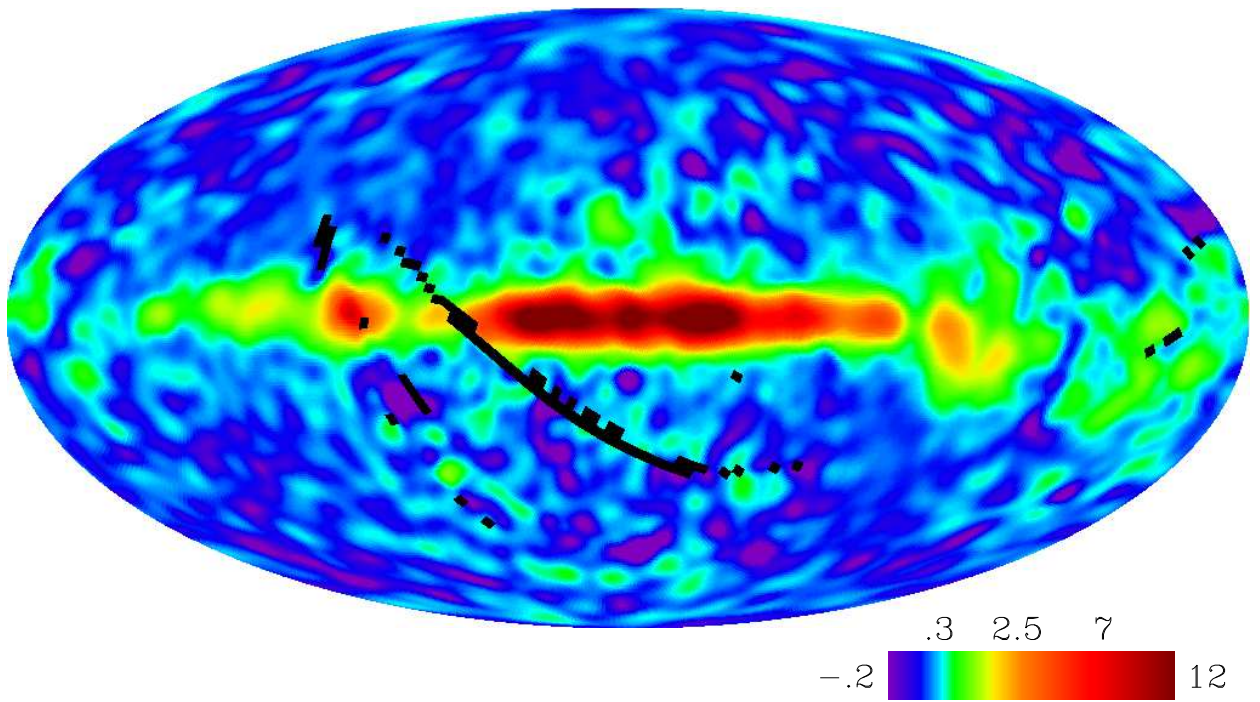
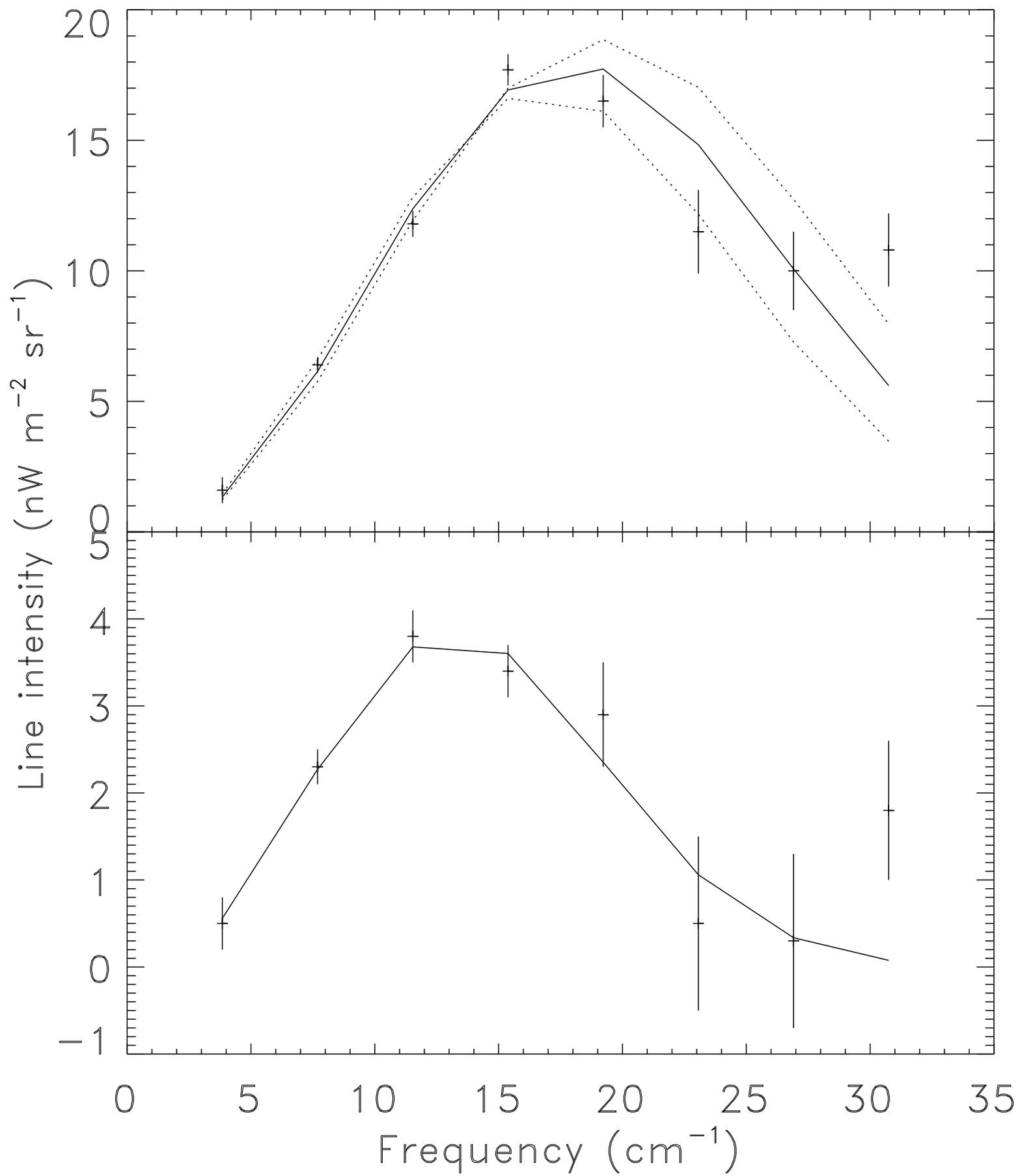
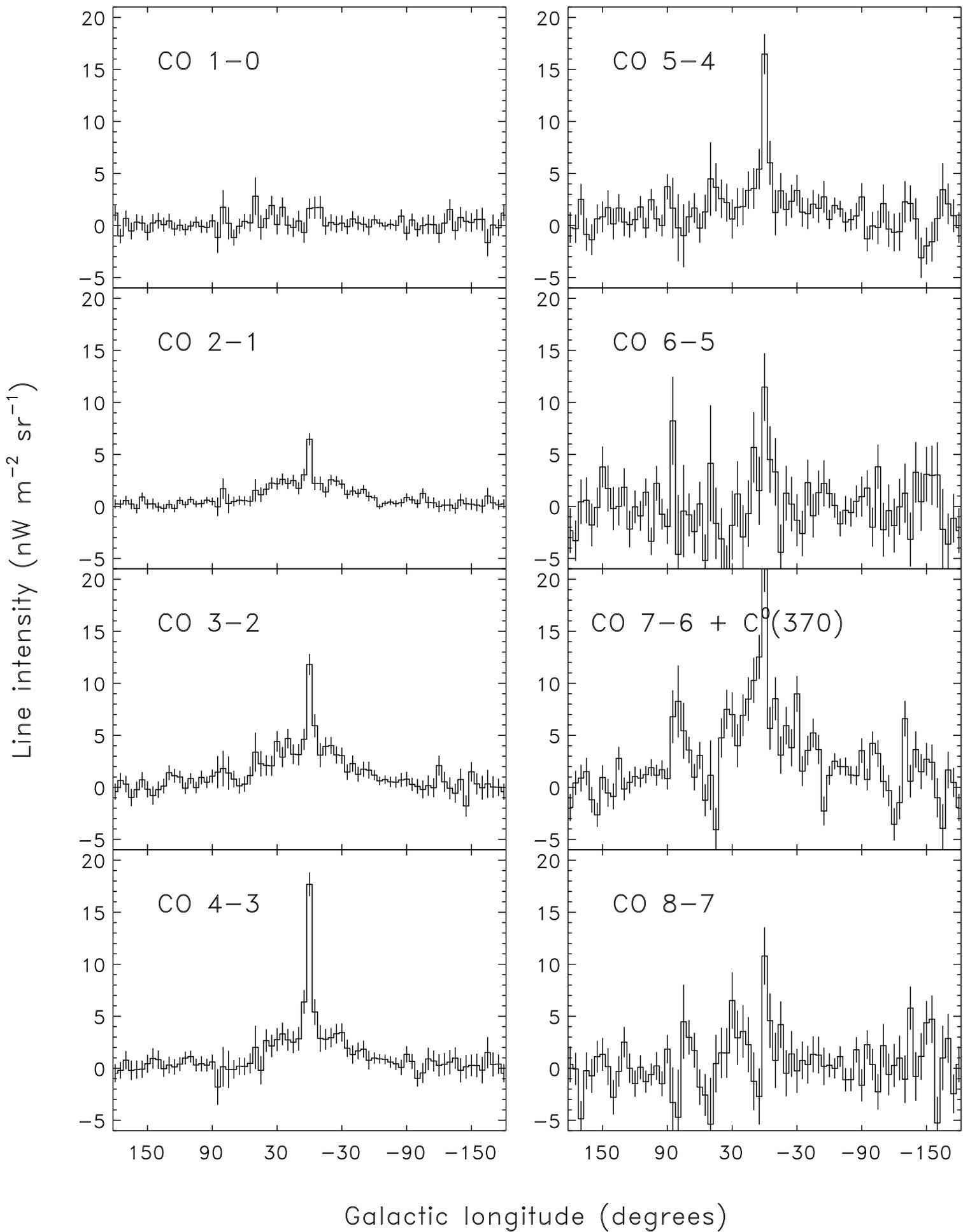
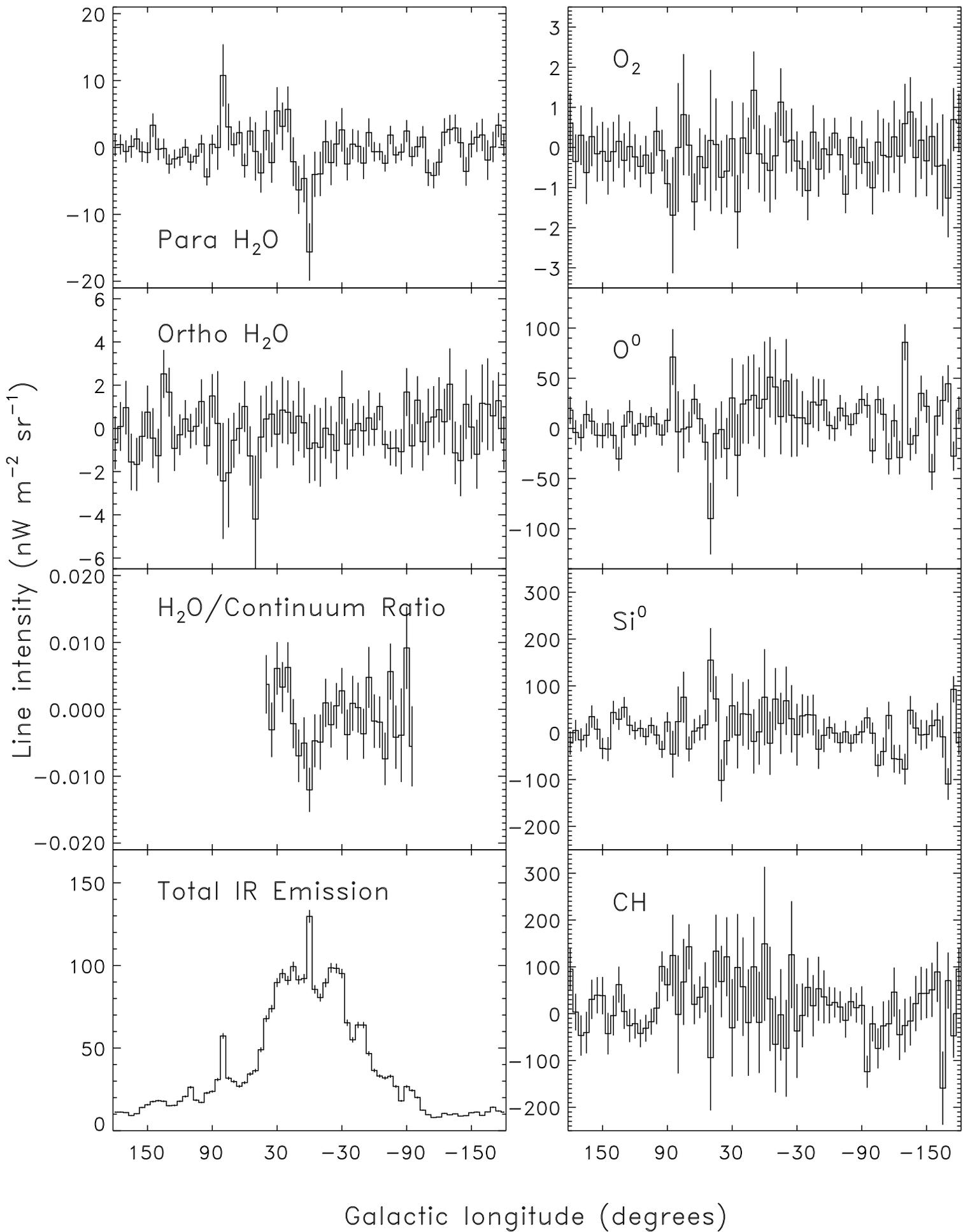
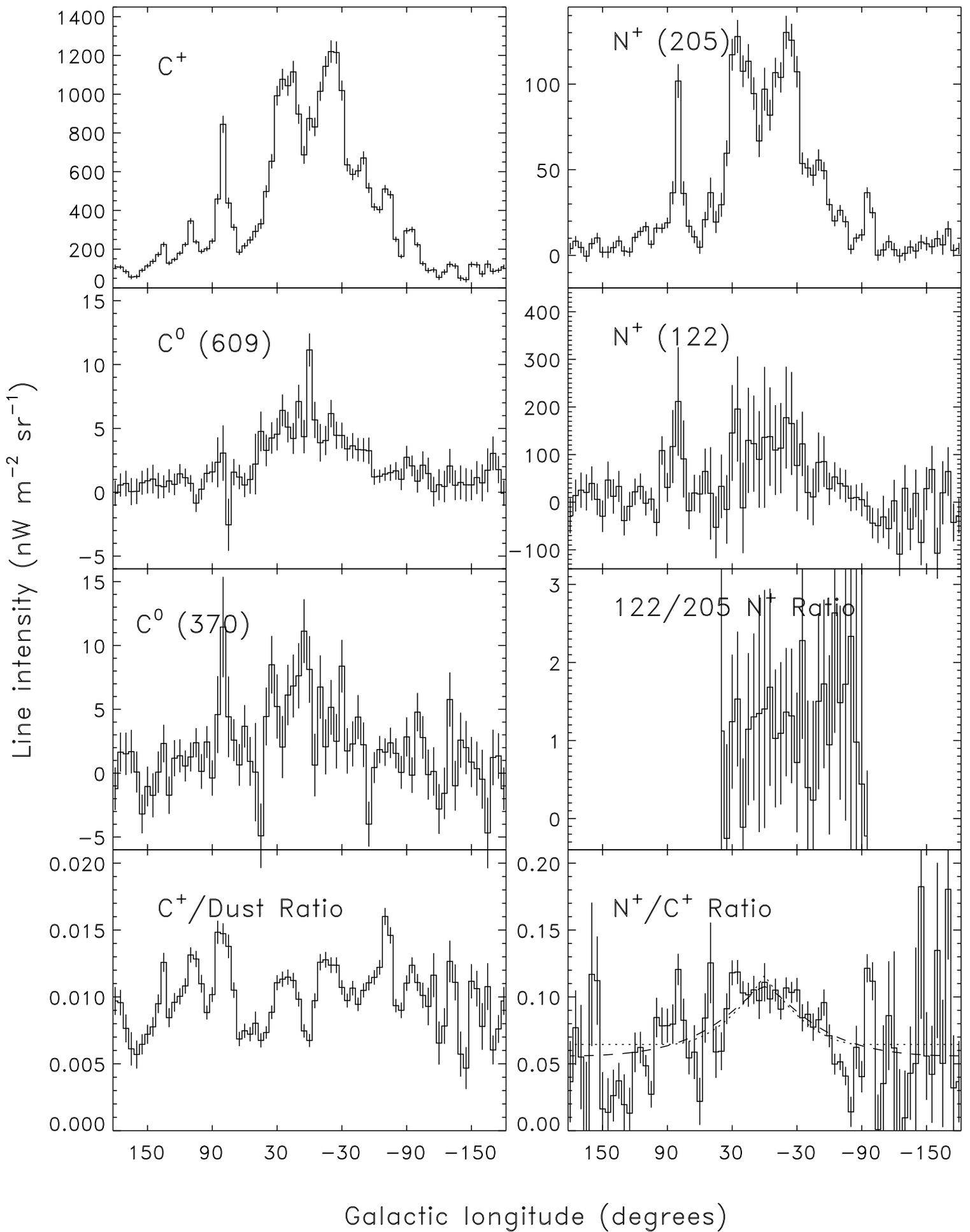
\end{document}